\documentclass[aps,prb,twocolumn]{revtex4}
\usepackage{bm,amsmath,graphicx}
\begin{document}

\title{General recursive solution for one-dimensional quantum potentials: a simple tool for applied physics}

\author{S\'ergio L. Morelh\~ao}
\author{Andr\'e V. Perrotta}
\affiliation{Instituto de F\'{\i}sica, Universidade de S\~ao Paulo,
CP 66318, 05315-970 S\~aoPaulo, SP, Brazil}

\begin{abstract}
A revision of the recursive method proposed by S.A. Shakir [Am. J.
Phys. \textbf{52}, 845 (1984)] to solve bound eigenvalues of the
Schr\"odinger equation is presented. Equations are further
simplified and generalized for computing wave functions of any
given one-dimensional potential, providing accurate solutions not
only for bound states but also for scattering and resonant states,
as demonstrated here for a few examples.
\end{abstract}

\maketitle

\section{Introduction}

Predicting physical properties of quantum systems has basically
been treated as an eigenvalue problem. Since from single atoms and
molecules up to nanostructured devices, chemical and electronic
properties are determined by finding the eigenstates of the
system's Hamiltonian. Methods of solving eigenvalue/eigenfunction
problems are therefore an inevitable part of the quantum mechanic
theory. In real systems, even one dimensional problems can be quite
difficult to solve analytically and often approximative, such as in
perturbation theory, and/or numerical methods are
required.\cite{hart1957, bohm1989, lent1990, lee_1991, jang1995,
simo1997} Semiconductor heterostructures are good examples of one
dimensional systems of technological interest where numerical
solutions of the Schr\"odinger equation (SE) are often employed for
self-consistent studies of electronic devices.\cite{kluk1989,
fren1990, timo2004, abda2006, mehd2006} Studies of vibrational
bound states of molecules in physical chemistry, as well as of
other oscillatory systems in atomic and nuclear physics are treated
as one dimensional problems when summarized in the radial SE; they
consist another class of important examples in which numerical
methods are also often employed. \cite{simo1996, simo2001,
vyve2005, chin2006}

Although many procedures and algorithms are available in nowadays
for solving the SE numerically \textemdash see for instance Refs.
\onlinecite{ledo2005}, \onlinecite{moye2004}, and references
therein\textemdash\, they are suitable to address specific problems
on distinct fields of applied and theoretical physics. There is
still a need for a general treatment that can be summarized into a
single routine capable to solve equally well problems of confined
modes by arbitrary potential as well as scattering problems in open
systems. Moreover, all available numerical methods are essentially
mathematical solutions of a differential equation, and hence the
level of expertise required for dealing with these methods has kept
them out of most physics and chemistry classrooms, even on graduate
courses, where the demonstrative examples are still limited to a
few cases of quantum potentials for which the analytical solutions
are known.

More than 20 years ago, S.A. Shakir \cite{shak1984} presented a
recursive method to find bound eigenvalues of the SE. It was a
relatively simple method, closely related to those used in the
field of optics for calculating the Fresnel amplitudes. In despite
of its simplicity, and even of the fact of have been rediscovered a
few years latter,\cite{kalo1991a,kalo1991b,senn1992} this method
has passed unnoted from the applied research fields until very
recently.\cite{bisw2005}

In this work, instead of concerning with mathematical methods to
solve the SE for bound and scattering states, we are concerned with
alternative methods capable of providing better physical insight
about the quantum world. In other words, methods that provide
solutions of the SE without having to solve it mathematically. One
class of alternative methods, which we call physical methods, is
possible in one-dimension. They are based on the elemental action
of a given stationary force field on the particle's wave function.
Action that occurs at every step of a discretized potential energy
function and it is qualitatively always the same, reflecting and
transmitting the incident wave as in the Shakir's
method.\cite{shak1984} However, here, the original formalism is
simplified and generalized into a procedure for computing wave
functions of any given potential whatever related to scattering,
resonance, tunneling, or bound states, as demonstrated for a few
examples. Physical method at their actual stage of development
might require higher computational cost than mathematical methods,
but offers versatility and exactness in return, as well as the fact
that they can be explained and used even by undergraduate students
and no-experts in numerical methods.

\section{Theory}

For a given potential energy function $U(x)$, it is always possible
to define the interval ${\bf X} = [x_0, x_N]$ such that outside its
boundaries, i.e. for $x \notin {\bf X}$, either the potential is
constant or the amplitude of the wave function vanish before
appreciable variation of the potential. In this interval, the
potential has been discretized into $N$ steps, and we seek for
stationary states with wave functions in the form
\begin{equation}
\Psi(x) = \sum_{j=0}^N\delta_j(x)\underrightarrow{\psi}_j(x)
\label{psi(x)}
\end{equation}
in which
\begin{equation}
\underrightarrow{\psi}_j(x)=A_je^{ik_j(x-x_j)}+B_je^{-ik_j(x-x_j)}
\label{psijeq}
\end{equation}
and $\delta_j(x)=1$ for $x \in [x_j,x_{j+1}]$ and $0$ otherwise.
The under arrow indicates that this format will be used here for
left-hand solutions, as better explained below.

Incremental distances ${\rm d}x_j = x_{j+1}-x_j$, are small enough
to guarantee a good numerical solution when taking
$U(x)=U(x_j)\delta_j(x)$, as shown in Fig. 1, so that
\begin{equation}
k_j = \varphi\sqrt{E-U(x_j)} \label{kj(E)}
\end{equation}
where $\varphi = \sqrt{2m/\hbar^2}$. $E$ and $m$ stand for energy
and mass of the particle, respectively.

\begin{figure}
\includegraphics[width=3.2in]{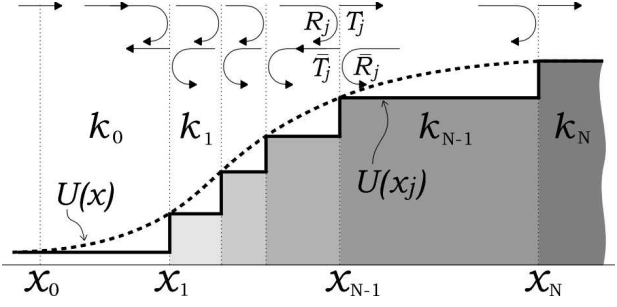}
\caption{Discretization of a energy potential function $U(x)$
(dashed line), providing reflection, $R_j$ and ${\bar R}_j$, and
transmission, $T_j$ and ${\bar T}_j$, coefficients at every step of
the discretized function (solid line). Coefficients with and
without bar stand for right-hand and left-hand scattering
solutions, respectively. Wavevectors $k_j$ at each interval of
constant potential value $U(x_j)$ is also shown.}
\end{figure}

To calculate the $A_j$ and $B_j$ amplitudes recursively, initial
values at the boundaries of the interval ${\bf X}$ have to be
provided. For instance, normalized solutions for incidence from the
left-hand side, i.e. incidence from $x<x_0$, must have $B_N = 0$
and $A_0 = 1$. Conservation of the probability current imposes that
\begin{equation}
\underrightarrow{\psi}_{j-1} = \underrightarrow{\psi}_j ~~{\rm
and}~~ \underrightarrow{\psi}_{j-1}' = \underrightarrow{\psi}_j'
\label{cceqs}
\end{equation}
at every step of the potential, i.e. at every $x_j$. The prime
symbol indicates the first derivative in $x$. Left-hand solutions
are obtained by applying these conditions, Eqs.~(\ref{cceqs}),
first at $x_N$ where $B_N=0$ and
\begin{equation}
A_N =
A_{N-1}\frac{2k_{N-1}}{k_{N-1}+k_N}e^{ik_{N-1}(x_N-x_{N-1})}=A_{N-1}T_N~,
\label{TNeq}
\end{equation}
and then at $x_{N-1}$ where
\begin{equation}
B_{N-1} =
A_{N-1}\frac{k_{N-1}-k_N}{k_{N-1}+k_N}e^{2ik_{N-1}(x_N-x_{N-1})} =
A_{N-1}R_N~ \label{RNeq}
\end{equation}
and
\begin{widetext}
\begin{equation}
A_{N-1} = A_{N-2}\frac{2k_{N-2}}{(k_{N-2} - k_{N-1})R_N +
(k_{N-2}+k_{N-1})}e^{ik_{N-2}(x_{N-1}-x_{N-2})}=A_{N-2}T_{N-1}~,
\label{TN-1eq}
\end{equation}
and then at $x_{N-2}$ where
\begin{equation}
B_{N-2} = A_{N-2}\frac{(k_{N-2}+k_{N-1})R_N +
(k_{N-2}-k_{N-1})}{(k_{N-2}-k_{N-1})R_N + (k_{N-2}+k_{N-1})}
e^{2ik_{N-2}(x_{N-1}-x_{N-2})} = A_{N-2}R_{N-1} \label{RN-1eq}
\end{equation}
\end{widetext}
and $A_{N-2}$ is analogous to Eq.~(\ref{TN-1eq}) when replacing $N$
by $N-1$.

This is a recursive procedure, repeatable since $j=N$ down to $j=1$
as summarized by
\begin{equation}
A_j = A_{j-1}T_j ~~{\rm and}~~ B_j = A_jR_{j+1} \label{AjBj}
\end{equation}
where
\begin{subequations}
\begin{equation}
T_j = \frac{2k_{j-1}}{(k_{j-1}-k_j)R_{j+1} +
(k_{j-1}+k_j)}e^{ik_{j-1}(x_j-x_{j-1})} \label{Tjeq}
\end{equation}
and
\begin{equation}
R_j = \frac{(k_{j-1}+k_j)R_{j+1} +
(k_{j-1}-k_j)}{(k_{j-1}-k_j)R_{j+1} + (k_{j-1}+k_j)}
e^{2ik_{j-1}(x_j-x_{j-1})}. \label{Rjeq}
\end{equation}
\label{TjRjeq}
\end{subequations}
All amplitudes are proportional to $A_0$, and within the condition
assumed when defining the interval ${\bf X}$, $R_{N+1}~=~0$.

Solutions for incident particles from the right-hand side, i.e.
incidence from $x > x_N$, are obtained in analogous procedure, but
for the sake of simplification in the final format of the recursive
equations it is convenient to redefine the amplitudes as follow
\begin{equation}
A_j = C_{j+1}e^{-ik_j{\rm d}x_j} ~~{\rm and}~~ B_j =
D_{j+1}e^{+ik_j{\rm d}x_j}, \label{Cj+1Dj+1}
\end{equation}
and hence the wave function $\Psi(x)$ of the stationary states will
be calculated by using
\begin{equation}
\underleftarrow{\psi}_j(x)=C_{j+1}e^{ik_j(x-x_{j+1})}+
D_{j+1}e^{-ik_j(x-x_{j+1})} \label{phijeq}
\end{equation}
instead of $\underrightarrow{\psi}_j(x)$.

By applying the continuity of $\underleftarrow{\psi}_j$ and
$\underleftarrow{\psi}'_j$, as in Eqs.~(\ref{cceqs}), analogous
deduction of that in Eqs.~(\ref{TNeq})-(\ref{RN-1eq}) can be
carried out but now starting from $x_1$ where $C_1 = 0$ up to
$x_{N+1}$ where $D_{N+1} = 1$ (for normalized solutions). It leads
to the recursive equations
\begin{equation}
D_j = D_{j+1}{\bar T}_j ~~{\rm and}~~ C_j = D_j {\bar R}_{j-1}
\label{DjCj}
\end{equation}
of the right-hand solutions where
\begin{subequations}
\begin{equation}
{\bar T}_j = \frac{2k_j}{(k_j - k_{j-1}){\bar R}_{j-1} + (k_j +
k_{j-1})}e^{ik_j(x_{j+1}-x_j)}~ \label{Tpjeq}
\end{equation}
and
\begin{equation}
{\bar R}_j = \frac{(k_j + k_{j-1}){\bar R}_{j-1} + (k_j -
k_{j-1})}{(k_j - k_{j-1}){\bar R}_{j-1} + (k_j + k_{j-1})}
e^{2ik_j(x_{j+1}-x_j)} \label{Rpjeq}
\end{equation}
\label{TpjRpjeq}
\end{subequations}
for $j = 1,2,\ldots,N$ and ${\bar R}_0=0$.

At certain energies $E=\mathcal{E}_{\nu}$, quantum confinements can
occur at local minima of the potential in which
$U(x)<\mathcal{E}_{\nu}$; given that each minimum is bounded on
both sides by non tunnelable barriers. Around one of such local
minima, stationary wave solutions only exist when the waves
reflected at both bounding ``walls'' interfere constructively to
each other at any instant of time. It means that, left-hand and
right-hand scattering solutions must coexist in the confinement
region, i.e. $\underrightarrow{\psi}_j(x) =
\underleftarrow{\psi}_j(x)$, which takes us back to
Eqs.~(\ref{Cj+1Dj+1}). Since $B_j = A_jR_{j+1}$ and $C_{j+1} =
D_{j+1}{\bar R}_j$, as given by Eqs.~(\ref{AjBj}) and (\ref{DjCj}),
all amplitudes cancel each other out in Eq.~(\ref{Cj+1Dj+1}) so
that ${\bar R}_jR_{j+1} = \exp(2ik_j{\rm d}x_j)$. This latter
relationship provides a criterion for numerical determination of
the allowed modes in any confinement region of the potential since
a minimization function such as
\begin{equation}
f(E) = \sum_j\left|{\bar R}_jR_{j+1} - e^{2ik_j{\rm d}x_j}\right|
\label{f(En)}
\end{equation}
can be defined with $j$ running over all values where $U(x_j)<E$.
Then, the condition $f(E) = 0$ is fulfilled only for a set of
discrete $\mathcal{E}_{\nu}$ values of energy, corresponding to the
eigenvalues of the one dimensional SE at a given local minimum of
$U(x)$.

Eigenfunctions $\Psi_{\nu}$, of the energy eigenvalues
$\mathcal{E}_{\nu}$, are computed as either
$\underrightarrow{\psi}_j(x)$ or $\underleftarrow{\psi}_j(x)$
solutions in the classical allowed region around the minimum. But
both solutions are required to compute the evanescent parts of the
eigenfunctions inside the bounding walls, and then it is necessary
to match the amplitudes of these solutions at some point. If the
$h$th step, for which $U(x_h)<\mathcal{E}_{\nu}$, is taken as the
matching point, $\underrightarrow{\psi}_h =
\underleftarrow{\psi}_h$ and hence $D_h = A_h(1+R_{h+1})/(1+{\bar
R}_{h-1})$. By providing an arbitrary value to $A_h$, such as
$A_h=1$, the non-normalized eigenfunction

\begin{equation}
\Psi_{\nu}(x) = \sum_j\delta_j(x)\times \left \{
  \begin{array}{cc}
  \underrightarrow{\psi}_j(x), & {\rm if}~h \leq j \leq N \\
  \underleftarrow{\psi}_j(x), & {\rm if}~1 \leq j < h ~
 \end{array}
\right. \label{psitotal}
\end{equation}
can be calculated over the entire interval ${\bf X}$. Since $S =
\sum_{j=0}^{N+1}|\Psi_{\nu}(x_j)|^2{\rm d}x_j$ has to be equal to
unit for a normalized eigenfunction, $\sqrt{S}$ is the
normalization factor.

\section{Examples and discussions}

A simple application of this recursive method is in finding
transmission coefficients,
\begin{equation}
{\rm T} = |A_N/A_0|^2 = 1 - |B_0/A_0|^2, \label{Tcoeff}
\end{equation}
across arbitrary potential barriers. It is known as being able to
provide exact results even for truncated potentials where the most
common approximative methods are used to fail, as compared
elsewhere.\cite{zhan2000} Besides exactness, another advantage of
the present formalism is that it is not just computing the values
of T, but the entire wave function instead. Hence, for more complex
potential barriers, as those usually find in heterostructures and
nanodevices where resonance can take place inside the
barries,\cite{trib2002,zhu1994,moye2004,bisw2005} amplitudes and
evanescence times of resonant modes are simultaneously computed.

\begin{figure}
\includegraphics[width=3.2in]{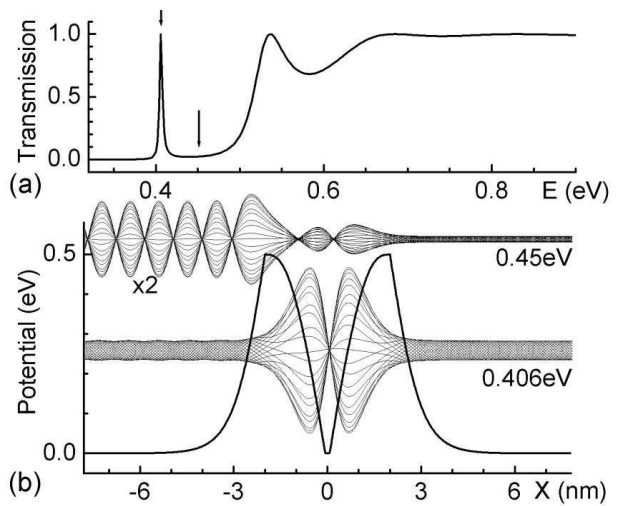}
\caption{Resonant tunneling in a double barrier with V-shaped well
at the middle. (a) Transmission coefficient T, Eq.~(\ref{Tcoeff}),
as a function of the particle's energy $E$; plot resolution is
$dE=0.002$eV. (b) Potential energy function of the barrier and wave
functions for the two energies (0.406eV and 0.45eV) pointed out by
arrows in (a). A set of curves is shown for each wave function
solution, corresponding to $\Psi(x)e^{-i\omega t}$ for several
values of $\omega t$. The amplitude of the wave function for
$E=0.45$eV has been magnified by a factor of 2 regarding the
amplitude of the resonant wave. Vertical displacements of both sets
of curves are not related to the energy scale at left. Solutions
computed for $m=511{\rm keV/c}^2$, i.e. $\varphi = 5.1232{\rm
eV}^{-1/2}{\rm nm}^{-1}$ in Eq.~(\ref{kj(E)}), and an uniform step
width of $dx_j = 0.02$nm in the interval $x_{N,0}=\pm10$nm
(N=1000).}
\end{figure}

As an example, consider the potential given in Fig. 2, which is a
double barrier with V-shaped well the middle. By scanning the
energy values $E$, Eq.~(\ref{kj(E)}), in a chosen range from
$E_{min}$ to $E_{max}$ in $N_E$ steps of width $dE$ so that $E_n =
E_{min} + (n-1)dE$ and $E_{N_E}=E_{max}$, a T($E$) curve is
obtained as the one shown in Fig. 2(a). By storing the coefficients
$A_j(E_n)$ and $B_j(E_n)$ when calculating the T values,
high-resolution plots of the wave functions in the interval ${\bf
X}$ also as a function of energy are already available, i.e.
\begin{equation}
\Psi(x,E_n) \simeq \sum_{j=0}^N \delta_j(x)[
A_j(E_n)+B_j(E_n)],\label{psi(x,En)}
\end{equation}
given that $dx_j << \lambda_j = 2\pi/|k_j|$. Hence, amplitudes and
attenuations of the waves through the entire barrier can be
visualized. For this particular potential, it shows that the mode
undergoing resonant-tunneling, at $E=0.406$eV, has two maxima in
the well as can be seen in Fig. 2(b).

Optionally one can compute only the $R_j$'s and $T_j$'s,
Eq.~(\ref{TjRjeq}), to obtain ${\rm T}=|{\small\prod}_{j=1}^N
T_j|^2$ and either stop the calculation there if only the $T(E)$
curve is desired or compute the $A_j$'s and $B_j$'s,
Eqs.~(\ref{AjBj}), later for a chosen set of energies. The options
on how the present formalism can be programmed for optimizing
computational tasks are detailed in Appendix A.

Although the stored set of solutions, Eq.~(\ref{psi(x,En)}), could
be used to study the scattering of wave packets in the barriers,
obtaining normalized wave packets would not be as straightforward
as possible because the solutions are evenly spaced in energy, not
in momentum. Therefore, if one wants to compute wave solutions only
once, and also study wave packets composed of these solutions as
well as their transmission coefficients, it is better to first
define the form of the initial wave packet.

Gaussian wave packets has become a benchmark in computing time
evolution of this sort,\cite{moye2004} hence let use
\begin{equation}
\Psi(x,0) = \frac{1}{\sqrt{\sigma_x\sqrt{\pi}}}
e^{i\kappa_0x}e^{-(x-x_0)^2/2\sigma_x^2} \label{gwpx}
\end{equation}
as reference for our initial free-particle wave packet in $t=0$,
centered at $x_0$, and described by a single mode of energy $E_0$
so that $\kappa_0 = \varphi\sqrt{E_0}$. It is normalized since
$\int|\Psi(x,0)|^2dx=1$, and its full width at half maximum (FWHM),
$W$, can be chosen by providing
$\sigma_x~=~W/\sqrt{2\ln4}~\simeq~0.6 W$.

Numerov-type of solutions of the time-dependent SE can handled the
time evolution of such wave packet, Eq.~(\ref{gwpx}), through a
given potential barrier,\cite{moye2004} but in the present
recursive method it is not possible to start with such expression
of $\Psi(x,0)$ since it is an artificial expression that happens to
have the same shape, in $t=0$, of the actual wave packet
\begin{equation}
\Psi(x,t) =
\frac{1}{\sqrt{2\pi}}\sum_{n=1}^{N_E}c_n\Psi(x,E_n)e^{-iE_nt/\hbar}
\label{Psi(x,t)}
\end{equation}
where
\begin{equation}
c_n = \frac{1}{\sqrt{\sigma_k\sqrt{\pi}}}
e^{-(\kappa_n-\kappa_0)^2/2\sigma_k^2} \label{cn}
\end{equation}
and $\sigma_k = \sigma_x^{-1}.$ For sake of normalization, the wave
vectors $\kappa_n$ must be equally spaced, i.e.
$\kappa_{n+1}-\kappa_n = d\kappa$, so that
\begin{equation}
\sum_{n=1}^{N_E}|c_n|^2d\kappa = 1.\label{ncn}
\end{equation}
Since the $c_n$'s out of the range
$\kappa_n-\kappa_0=\pm3.5\sigma_k$ are of negligible amplitude, the
$N_E$ modes with wave vectors
$\kappa_1,\,\kappa_2,\ldots,\,\kappa_{N_E}$ are taken in this range
to compose the wave packet, whose spectrum of energy is given by
$E_n = (\kappa_n/\varphi)^2$. Energy difference between adjacent
modes, $$dE_n~=~E_{n+1}-E_n~\simeq~2\kappa_nd\kappa/\varphi^2,$$
increases as $n$ runs toward $N_E$. Hence, the maximum interval of
time by which the evolution of the wave packet can be studied is
\begin{equation}
\Delta t~\leq~\frac{h}{2(E_{N_E}-E_{N_E-1})} \approx
(N_E-1)\frac{h}{4\Delta E}\left(1-\frac{\Delta
E}{2E_0}\right)\label{Dt}
\end{equation}
when the energy range $E_0\pm\Delta E$ comprises all $E_n$'s.

\begin{figure}
\includegraphics[width=3.2in]{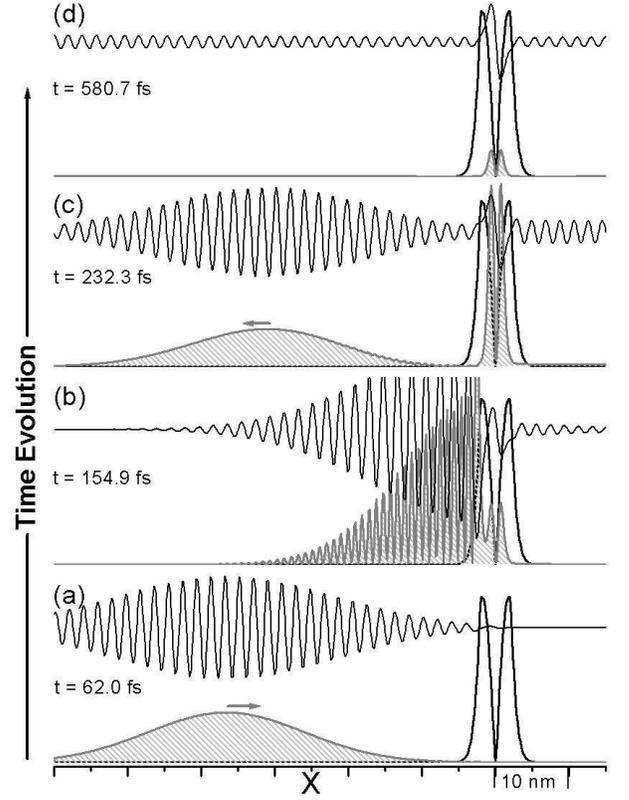}
\caption{Scattering of a gaussian wave packet in a double barrier
with V-shaped well (thick line, Fig.2). Time evolution of
$\Psi(x,t)$ (thin line) and $|\Psi(x,t)|^2$ (gray shaded curves),
increasing from (a) to (d), as indicated aside of each snapshot.
Wave packet obtained according to Eqs.~(\ref{Psi(x,t)}) and
(\ref{cn}), as described in the text. Group velocity $v_G =
0.378$nm/fs ($\kappa_0 = 3.264{\rm nm}^{-1}$), initial position at
$x_0=-60$nm from the center of the barriers, and FWHM of $25$nm
($\sigma_x = 1/\sigma_k = 15$nm). }
\end{figure}

To be more illustrative, let construct an initial guassian wave
packet within a previously chosen range of energy, for instance,
with $E_n$ in the narrow range $E_0\pm0.058$eV around the energy
$E_0=0.406$eV of the resonant mode in Fig. 2. It leads to $\sigma_k
\simeq \varphi\Delta E /7\sqrt{E_0}=0.0666{\rm nm}^{-1}$ and to $W
= 25.0$nm as the FWHM of $|\Psi(x,0)|^2$. By composing the wave
packet with 101 modes, our limit of time for studying its evolution
is $\Delta t \leq 1654$fs, Eq.~(\ref{Dt}). Fig. 3 shows the
scattering process of such wave packet by the double barrier (Fig.
2) as computed via Eq.~(\ref{Psi(x,t)}). The reflection of the
entire wave packet occurs in a time interval no longer than 170fs,
Fig. 3(a) through Fig. 3(c), but the resonant mode trapped in the
well survives for much longer than that, Fig. 3(d). Its evanescence
in time is dictated by an exponential decay of time constant of
$184.5$fs regarding the probability of finding the particle inside
the well after $t=232.3$fs, Fig. 3(c).

\begin{figure}
\includegraphics[width=3.2in]{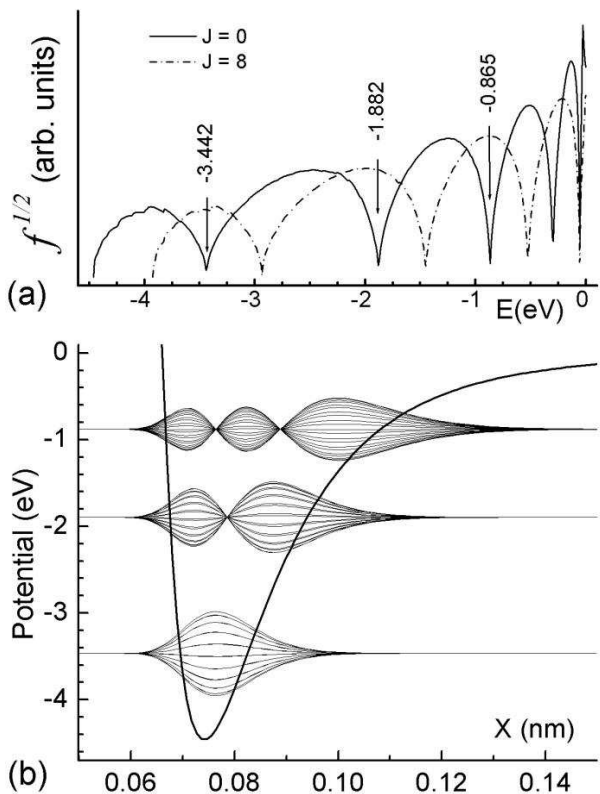}
\caption{(a) Vibrational energy eigenvalues estimated via
minimization of $f(E)$, Eq.~(\ref{f(En)}), for a Lennard-Jones type
of potential and for two rotational states, with $J=0$ (solid line)
and with $J=8$ (dash-dot line). Step width $dx_j = 0.95$pm.
Potential energy functions given by ${\nobreak U(x) =
A/x^{12}-B/x^{6} + J(J+1)/(\varphi x)^2}$ where $A =
0.124\times10^{-12}{\rm eV}\cdot{\rm nm}^{12}$, $B =
1.488\times10^{-6}{\rm eV}\cdot{\rm nm}^{6}$ and
$m=469.4$MeV/c$^2$. (b) Eigenfunctions of the three first
vibrational states for the case $J=0$, indicated by arrows in (a),
are shown in the same scheme as before (Fig. 2), i.e. in sets of 20
curves per time period of amplitude oscillation. Wave amplitudes
are in a common scale. Plot of $U(x)$ (thick-solid line), $J=0$, is
also shown.}
\end{figure}

In case of potential with localized minima not reachable by
tunneling from either sides, computing both left-hand and
right-hand solutions is necessary. The procedure is very similar to
the previous one used to obtain the $\Psi(x,En)$ set of wave
functions evenly spaced in energy where $E_{n+1}-E_n=dE$. But, most
of these unilateral solutions do not exist under confinement.
Hence, prior to calculate $A_j$'s, $B_j$'s, and $T_j$'s from
Eqs.~(\ref{AjBj}) and (\ref{Tjeq}), we first compute the $R_j$'s
and $\bar{R}_j$'s, Eqs.~(\ref{Rjeq}) and (\ref{Rpjeq}), to generate
the $f(E)$ plot according to Eq.~(\ref{f(En)}). In Fig. 4(a) there
are examples of such plot for a Lennard-Jones potential, a type of
potential used to describe diatomic molecules and that has a single
minimum. Only after providing an energy eigenvalue,
$\mathcal{E}_{\nu}$, all other required coefficients for plotting
the eigenfunction $\Psi_{\nu}(x)$, Eq.~(\ref{psitotal}), are
calculated. For $\nu = 1$, 2, and 3, the eigenfunctions are shown
in Fig. 4(b) where the one-dimensional variable $x$ stands for the
nuclei distance in the molecule. Although reduced mass, equilibrium
distance, and ionization energy values are close of those for the
H$_2$ molecule, the width of the depression in $U(x)$ is much
narrow than in the actual molecule.\cite{syrk1950} It means a
stronger restoring force and hence larger energy gaps in between
adjacent vibrational states, which is used here for illustrative
purposes only. In this hypothetical molecule, the non-constancy of
the gap values is more evident, as can be seen in Fig. 4(a). It
indicates how the exact solution would differ from the harmonic
oscillator approximation.

Computational task involved in finding the energy eigenvalues
$E=\mathcal{E}_{\nu}$ in which $f(E)=0$ can be enormous depending
on number and accuracy of the desired eigenvalues. There will be no
difficult in determining a few eigenvalues with reasonable
accuracy, for instance $\mathcal{E}_1=-3.4416$eV,
$\mathcal{E}_2=-1.8793$eV, $\mathcal{E}_3=-0.8660$eV, and
$\mathcal{E}_4=-0.2991$eV in Fig. 4(a), are obtained at once in
just a few seconds and with an accuracy of $dE/2 = 0.0022$eV
($N_E=1000$); $x_0=0.002$nm, $x_N=0.2$nm, and $dx_j = 0.0005$nm
($N=396$). The problem relies in finding many eigenvalues, with
high accuracy, occurring at different densities along a broad
energy range. But, practical needs of finding all possible
eigenvalues are unusual, although it can be realized by developing
search routines to reduce computational time since the method does
not have any intrinsic limitation; as for instance the frequency
distortion, or phase-lag, errors of the Numerov-type
methods.\cite{simo1996, simo2001}

Potential functions with singularities do not compromise the
applicability of the presented method since only solutions with
null wave functions at the singularities are those with physical
meaning. In other words, when representing the potential by a
discretized function, a finite value has to be assigned to the
singularity. It compromises mostly the modes with non-null wave
functions at the singularity, but these are artificial modes that
do not exist in the actual physical system. Occurrence of
artificial modes has been observed when solving the potential $U(x)
= -e^2/[|x|+|\epsilon|]$ with $\epsilon \rightarrow 0$. In this
case, the odd solutions with respect to the singularity provide the
well-known eigenvalues of the hydrogen atom.

Another situation that can also be addressed by the recursive
method occurs when the potential function has closely spaced
minima, as for instance in the potential shown in Fig. 5. In the
present formalism, the two minima (or wells) of this potential can
be treated separately after establishing distinct intervals in
$\bf{X}$ for calculating the function $f(E)$. For each
$E_n\in[E_{min},E_{max}]$, we first compute the $R_j$'s and
$\bar{R}_j$'s in the entire interval $\bf{X}$ that contains the
wells, i.e. from $x_0$ to $x_N$. Then, by selecting a position
$x_b$ somewhere in between the wells as indicated in Fig.~5(b), two
$f(E_n)$ values are obtained: one for $x_j\in[x_0,x_b]$ and another
for $x_j\in[x_{b+1},x_N]$, and always obeying that $U(x_j)<E_n$.
Both $f(E)$ curves thus obtained are shown in Fig.~5(a). As
tunneling across the barrier separating the wells become relevant,
the minima of the $f(E)$ curves are coincidental. It means that the
energy levels, or eigenvalues $\mathcal{E}_{\nu}$, in each well are
independent from each other only if tunneling is negligible. It is
more evident when visualizing the wave functions of the confined
modes shown in Figs.~5(b), 5(c), and (d). In practice, potential
with two minima are found for Cooper pairs (2 spin-coupled
electrons) in josephson-junction of superconducting
materials.\cite{jose1974}

\begin{figure}
\includegraphics[width=3.2in]{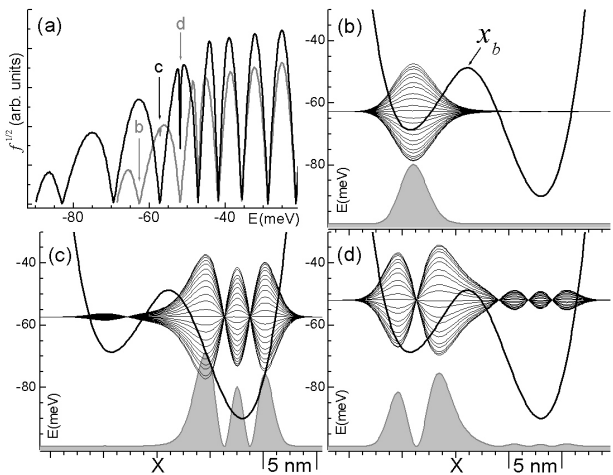} \caption{Energies and stationary wave functions of confined modes in a potential with two minima. (a) $f(E)$ curves for the right (black line) and left (gray line) minima, showing the allowed energies in the wells. (b), (c), (d) Confined modes at the energies $-62.68(12)$meV, $-57.27(13)$meV, and $-51.86(12)$meV, as indicated by arrows in (a). $\Psi(x)e^{-i\omega t}$ (thin lines) for several values of $\omega t$, and $|\Psi(x)|^2$ (gray shaded curves) are shown as well as the plot of ${\nobreak U(x)=A(x-a)^2+B\exp[-(x-a-\delta)^2/\alpha^2]+C}$ (thick line); $A = 4.0$meV ($=2.4$meV) for $x\leq a$ ($x > a$), $B=450$meV, $C=-500$meV, $\delta = 0.5$nm, and $\alpha=10$nm. $x_{N,0}=a\pm20$nm and $dx_j=0.1$nm (N=400). Particle's mass $m = 1.022{\rm MeV/c}^2$.}
\end{figure}

In semiconductor heterostructures, the validity of this recursive
method in its present formalism holds true when the charge
carrier's effective mass can be approximated by a constant value
through out the structure. However, accounting for effective mass
variation across heterojunctions has only been possible in cases of
abrupt interfaces and by using appropriated boundary
conditions.\cite{bast1991,wang2004}

Computing wave functions recursively is limited to problems
treatable in one dimension, such as spherically symmetric
potentials, or cases where the potentials in orthogonal directions
are independent from each other, as for instance ${\nobreak
U(x,y)=U_X(x)+U_Y(y)}$, in which the wave functions are obtained by
the product of independent one-dimensional solutions. In all other
cases of two and three dimensions, it is not possible to extend
this recursive method since the number of unknown amplitude
coefficients are more than the number of equations written by
matching the wave function and its derivative at every steps of the
discretized potentials. Hence only mathematical methods for solving
2D and 3D differential equations are
available.\cite{lent1990,jang1995,hwan2004,wang2004}

\section{Conclusions}

This work has demonstrated that any one-dimensional quantum
potential, wherever representing an open or closed system, can be
solved by a single method and within good numerical accuracy. It is
a simple tool ready to be used in applied physics or just in
developing illustrative examples of quantum mechanics. It should
also be emphasized that, in the presented formulation, the quantum
potentials are solved by a physical treatment where the particle's
wave function is molded by the same elemental action of the given
force field. This is a completely different approach of the one
that is usually taken where, once the potential is known, the
problem became the mathematical solution of a differential
equation.

\appendix
\section{Routines for computing wave solutions recursively}

There are several ways to program the recursive formalism presented
in this work, depending on what information is desired. The
flowchart in Fig. 6 describes the basic structure of the routines
used to elaborate the examples in Figs. 2 to 5. Once the $U(x)$
function and the discretization of $\bf{X}$ is defined, the user
can choose either route \textcircled{\scriptsize 1} or
\textcircled{\scriptsize 2} to scan a given energy range in fixed
increments of energy, $dE$, or wavevector, $d\kappa$, respectively.
The latter route is suitable to study time evolution of wave
packets by ending through route \textcircled{\scriptsize 7}, i.e.
\textcircled{\scriptsize 2}$\rightarrow$\textcircled{\scriptsize
3}$\rightarrow$\textcircled{\scriptsize
6}$\rightarrow$\textcircled{\scriptsize 7}. Eigenvalues
$\{\mathcal{E}_1,\,\mathcal{E}_2,\ldots\}$, of confined modes are
obtained from the $f(E)$ curve in route \textcircled{\scriptsize
4}, and used as input to provide the eigenfunctions $\Psi_{\nu}(x)$
through route \textcircled{\scriptsize 8}. The straight route
\textcircled{\scriptsize 3}$\rightarrow$\textcircled{\scriptsize 5}
are exclusive for users interested only in transmission
coefficients across arbitrary potential barriers.

\begin{figure*}
\includegraphics[width=6.2in]{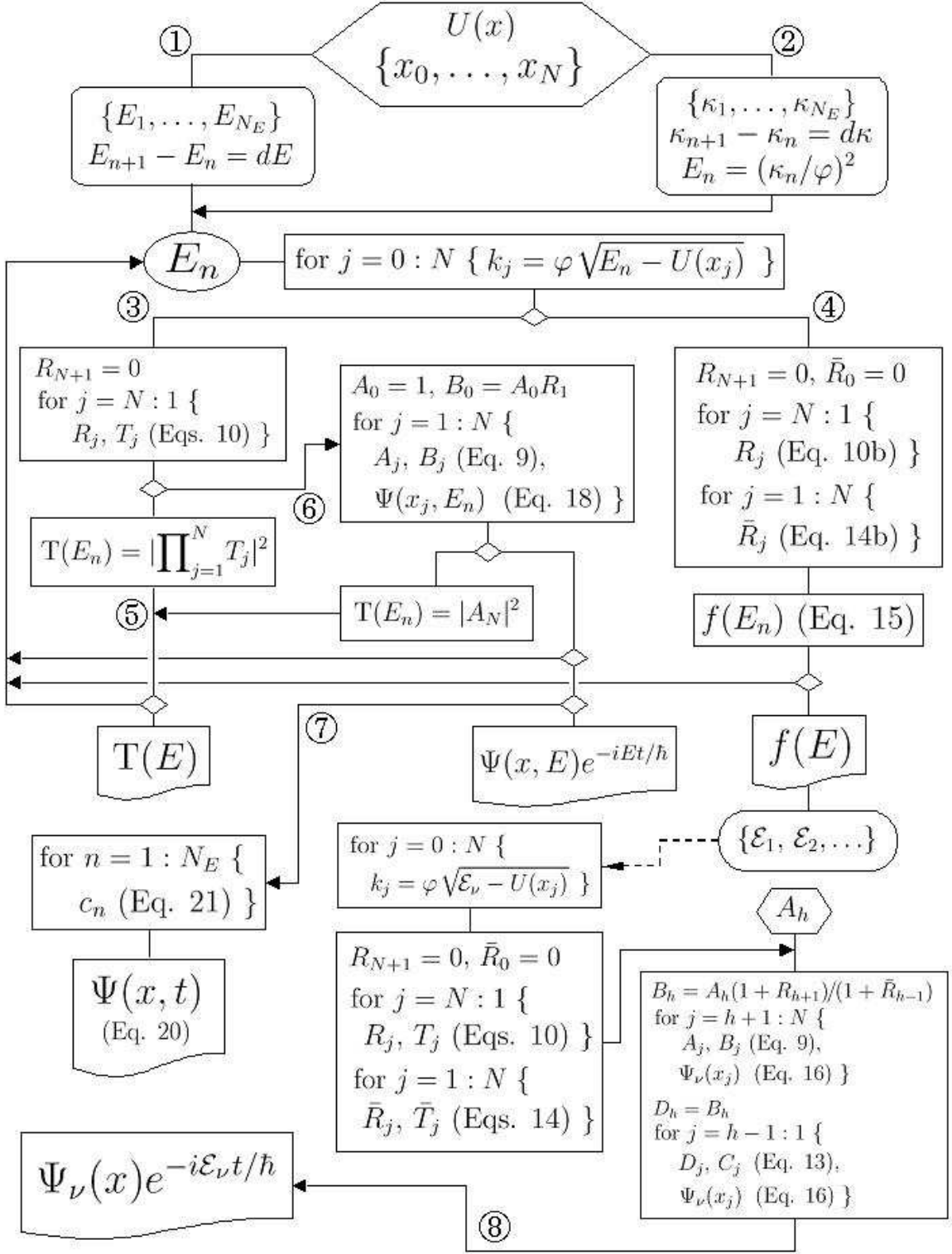}
\caption{Flowchart of possible routines for studying one
dimensional potentials with the present recursive formalism.
Transmission coefficients, $T(E)$ curves only:
\textcircled{\scriptsize 1}$\rightarrow$\textcircled{\scriptsize
3}$\rightarrow$\textcircled{\scriptsize 5}. Wave function
$\Psi(x,E)$ plots, and/or $T(E)$ curves: \textcircled{\scriptsize
1}$\rightarrow$\textcircled{\scriptsize
3}$\rightarrow$\textcircled{\scriptsize 6}. Time evolution of wave
packet $\Psi(x,t)$ plots: \textcircled{\scriptsize
2}$\rightarrow$\textcircled{\scriptsize
3}$\rightarrow$\textcircled{\scriptsize
6}$\rightarrow$\textcircled{\scriptsize 7}. Eigenvalues
$\mathcal{E}_{\nu}$, and eigenfunction $\Psi_{\nu}(x)$ plots:
\textcircled{\scriptsize 1}$\rightarrow$\textcircled{\scriptsize
4}$\rightarrow$\textcircled{\scriptsize 8}.}
\end{figure*}

\begin{acknowledgments}
During the time dedicated to this work, financial supports were
received from CNPq (301617/1995-3 and 304916/2006-4), FAPESP
(2002/10387-5), and CAPES-GRICES (140/05).
\end{acknowledgments}

\end{document}